\documentclass[twoside,twocolumn,9pt]{article}
\usepackage{extsizes}
\usepackage[super,sort&compress,comma]{natbib} 
\usepackage[version=3]{mhchem}
\usepackage[left=1.5cm, right=1.5cm, top=1.785cm, bottom=2.0cm]{geometry}
\usepackage{balance}
\usepackage{amsmath}
\usepackage{amssymb}
\usepackage{mathptmx}
\usepackage{sectsty}
\usepackage{graphicx} 
\usepackage{lastpage}
\usepackage{color}
\usepackage[format=plain,justification=justified,singlelinecheck=false,font={stretch=1.125,small,sf},labelfont=bf,labelsep=space]{caption}
\usepackage{float}
\usepackage{fancyhdr}
\usepackage{fnpos}
\usepackage[english]{babel}
\addto{\captionsenglish}{%
  
}
\usepackage{array}
\usepackage{droidsans}
\usepackage{charter}
\usepackage[T1]{fontenc}
\usepackage[usenames,dvipsnames]{xcolor}
\usepackage{setspace}
\usepackage[compact]{titlesec}
\usepackage{hyperref}
\usepackage{epstopdf}%This line makes .eps figures into .pdf - please comment out if not required.

\definecolor{cream}{RGB}{222,217,201}

\begin{document}

\pagestyle{fancy}
\thispagestyle{plain}
\fancypagestyle{plain}{
%%%HEADER%%%
\renewcommand{\headrulewidth}{0pt}
}
%%%END OF HEADER%%%

%%%PAGE SETUP - Please do not change any commands within this section%%%
\makeFNbottom
\makeatletter
\renewcommand\LARGE{\@setfontsize\LARGE{15pt}{17}}
\renewcommand\Large{\@setfontsize\Large{12pt}{14}}
\renewcommand\large{\@setfontsize\large{10pt}{12}}
\renewcommand\footnotesize{\@setfontsize\footnotesize{7pt}{10}}
\makeatother

\renewcommand{\thefootnote}{\fnsymbol{footnote}}
\renewcommand\footnoterule{\vspace*{1pt}% 
\color{cream}\hrule width 3.5in height 0.4pt \color{black}\vspace*{5pt}} 
\setcounter{secnumdepth}{5}

\makeatletter 
\renewcommand\@biblabel[1]{#1}            
\renewcommand\@makefntext[1]% 
{\noindent\makebox[0pt][r]{\@thefnmark\,}#1}
\makeatother 
\renewcommand{\figurename}{\small{Fig.}~}
\sectionfont{\sffamily\Large}
\subsectionfont{\normalsize}
\subsubsectionfont{\bf}
\setstretch{1.125} %In particular, please do not alter this line.
\setlength{\skip\footins}{0.8cm}
\setlength{\footnotesep}{0.25cm}
\setlength{\jot}{10pt}
\titlespacing*{\section}{0pt}{4pt}{4pt}
\titlespacing*{\subsection}{0pt}{15pt}{1pt}
%%%END OF PAGE SETUP%%%

%%%FOOTER%%%
\fancyfoot{}
\fancyfoot[LO,RE]{\vspace{-7.1pt}\includegraphics[height=9pt]{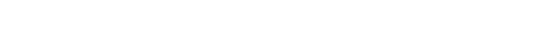}}
\fancyfoot[CO]{\vspace{-7.1pt}\hspace{13.2cm}\includegraphics{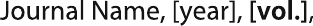}}
\fancyfoot[CE]{\vspace{-7.2pt}\hspace{-14.2cm}\includegraphics{head_foot/RF}}
\fancyfoot[RO]{\footnotesize{\sffamily{1--\pageref{LastPage} ~\textbar  \hspace{2pt}\thepage}}}
\fancyfoot[LE]{\footnotesize{\sffamily{\thepage~\textbar\hspace{3.45cm} 1--\pageref{LastPage}}}}
\fancyhead{}
\renewcommand{\headrulewidth}{0pt} 
\renewcommand{\footrulewidth}{0pt}
\setlength{\arrayrulewidth}{1pt}
\setlength{\columnsep}{6.5mm}
\setlength\bibsep{1pt}
%%%END OF FOOTER%%%

%%%FIGURE SETUP - please do not change any commands within this section%%%
\makeatletter 
\newlength{\figrulesep} 
\setlength{\figrulesep}{0.5\textfloatsep} 

\newcommand{\topfigrule}{\vspace*{-1pt}% 
\noindent{\color{cream}\rule[-\figrulesep]{\columnwidth}{1.5pt}} }

\newcommand{\botfigrule}{\vspace*{-2pt}% 
\noindent{\color{cream}\rule[\figrulesep]{\columnwidth}{1.5pt}} }

\newcommand{\dblfigrule}{\vspace*{-1pt}% 
\noindent{\color{cream}\rule[-\figrulesep]{\textwidth}{1.5pt}} }

\makeatother
%%%END OF FIGURE SETUP%%%

%%%TITLE, AUTHORS AND ABSTRACT%%%
\twocolumn[
  \begin{@twocolumnfalse}
{\includegraphics[height=30pt]{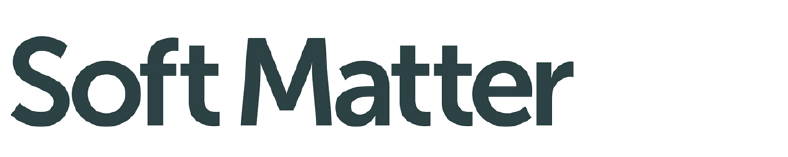}\hfill\raisebox{0pt}[0pt][0pt]{\includegraphics[height=55pt]{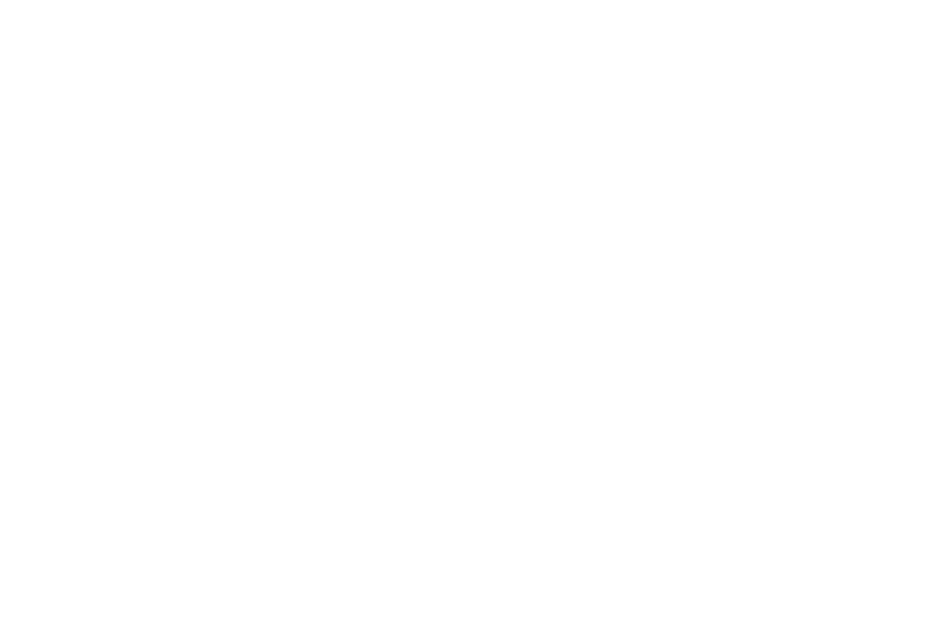}}\\[1ex]
\includegraphics[width=18.5cm]{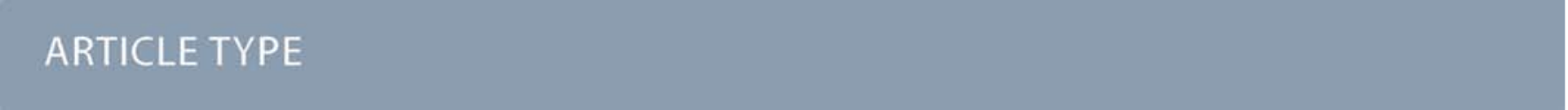}}\par
\vspace{1em}
\sffamily
\begin{tabular}{m{4.5cm} p{13.5cm} }

\includegraphics{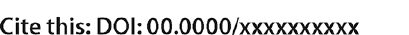} & \noindent\LARGE{\textbf{Rheology finds distinct glass and jamming transitions in emulsions$^\dag$}} \\%Article title goes here instead of the text "This is the title"
\vspace{0.3cm} & \vspace{0.3cm} \\

 & \noindent\large{Cong Cao,$^{\ast}$\textit{$^{a}$} Jianshan Liao,\textit{$^{b}$} Victor Breedveld,\textit{$^{b}$}and Eric R. Weeks$^{\ast}$\textit{$^{a}$}} \\%Author names go here instead of "Full name", etc.

\includegraphics{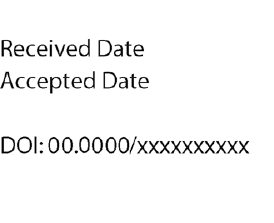} & \noindent\normalsize{We study the rheology of monodisperse and bidisperse emulsions with various droplet sizes (1 $\mu$m -- 2 $\mu$m diameter). Above a critical volume fraction $\phi_c$, these systems exhibit solid-like behavior and a yield stress can be detected.  Previous experiments suggest that for small thermal particles, rheology will see a glass transition at $\phi_c = \phi_g =0.58$; for large athermal systems, rheology will see a jamming transition at $\phi_c = \phi_J =0.64$.  However, simulations point out that at the crossover of thermal and athermal regimes, the glass and jamming transitions may both be observed in the same sample. Here we conduct an experiment by shearing four oil-in-water emulsions with a rheometer. We observe both a glass and a jamming transition for our smaller diameter droplets, and only a jamming transition for our larger diameter droplets.  The bidisperse sample behaves similarly to the small droplet sample, with two transitions observed. Our rheology data are well-fit by both the Herschel-Bulkley model and the Three Component model. Based on the fitting parameters,  our raw rheological data would not collapse onto a master curve. Our results suggest that liquid-solid transitions in dispersions may not be universal, but depend on particle type.} \\ 

\end{tabular}

 \end{@twocolumnfalse} \vspace{0.6cm}

  ]
%%%END OF TITLE, AUTHORS AND ABSTRACT%%%

%%%FONT SETUP - please do not change any commands within this section
\renewcommand*\rmdefault{bch}\normalfont\upshape
\rmfamily
\section*{}
\vspace{-1cm}

%%%FOOTNOTES%%%

{\footnotetext{\textit{$^{a}$Department of Physics, Emory University, Atlanta, GA 30322, USA; E-mail: cong.cao@emory.edu}}}
{\footnotetext{\textit{$^{b}$School of Chemical \& Biomolecular Engineering, Georgia Tech, Atlanta, GA 30332, USA }}}

%Please use \dag to cite the ESI in the main text of the article.
%If you article does not have ESI please remove the the \dag symbol from the title and the footnotetext below.
%\footnotetext{\dag~Electronic Supplementary Information (ESI) available: [details of any supplementary information available should be included here]. See DOI: 10.1039/cXsm00000x/}
%additional addresses can be cited as above using the lower-case letters, c, d, e... If all authors are from the same address, no letter is required

%\footnotetext{\ddag~Additional footnotes to the title and authors can be included \textit{e.g.}\ `Present address:' or `These authors contributed equally to this work' as above using the symbols: \ddag, \textsection, and \P. Please place the appropriate symbol next to the author's name and include a \texttt{\textbackslash footnotetext} entry in the the correct place in the list.}

%%%END OF FOOTNOTES%%%

%%%MAIN TEXT%%%%

\section{Introduction}

Soft amorphous solids include granular materials, foams, and pastes.  These are solid in the sense that they possess a yield stress:  they elastically (reversibly) support a finite stress below the yield stress, and deform irreversibly if the applied stress exceeds the yield stress \cite{bonn17}.  In particular, these materials can typically support their own weight:  you can make a pile of sand, a pile of shaving cream, or a pile of paste as well.  If you scoop mayonnaise out of a jar, the divot remains for a long time.  A granular material such as sand is comprised of solid particles, typically mm or larger sized.  A colloidal paste is made from solid particles in a liquid, typically $\mu$m size or smaller (small enough to undergo Brownian motion).  Foams are gas bubbles in a liquid, stabilized by surfactant molecules to prevent the bubbles from coalescing, and typically with mm or larger sized bubbles.  Another soft amorphous solid can be found in emulsions:  these are composed of liquid droplets in another immiscible liquid, again with surfactant molecules needed for stabilization, and with droplet sizes ranging from 10~nm up to a few hundred $\mu$m \cite{mason06}.  In all of these examples, by diluting the particles the material can lose its solid-like properties:  for example, adding water to a pile of shaving cream will eventually change the pile of foam into a puddle with bubbles.  More technically, the yield stress for these materials is a function of the volume fraction $\phi$:  as the fraction of particles in the volume is decreased, the material flows more easily.  For $\phi > \phi_c$, the yield stress becomes nonzero, with critical volume fraction $\phi_c$ depending on details of the system.  When these materials undergo the transition from liquid-like to solid-like behavior, these materials share some similarities with glass transition, where a liquid can be changed into an apparent amorphous solid by either increasing $\phi$ (equivalently, density) or decreasing temperature.\cite{andrea98}

Previously Liu, Nagel, and coworkers \cite{andrea98,ohern03,andrea10} presented the jamming framework to unify these transitions from liquid into solid.  They suggested that in order to change a jammed system into an unjammed one, there are three possible options: increasing the temperature, decreasing the volume fraction, and increasing the applied stress above the yield stress.  This can be restated as a conjecture that the yield stress is a universal function of temperature and volume fraction.  Focusing just on particulate systems such as the ones mentioned above, one would expect that granular materials, foams, colloids, and emulsions would share a common $\phi_c$ (at least if their particle size distributions are equivalent \cite{desmond14}).  However, it has long been noted that the ``colloidal glass transition'' happens at $\phi_g = 0.58$ \cite{pusey86,hunter12rpp}, and random close packing of granular particles happens at $\phi_{RCP}=0.63$ \cite{bernal64} (both situations considering essentially monodisperse hard particles).  For an emulsion, a 1995 rheology experiment by Mason, Bibette, and Weitz noted that there was evidence of solid-like behavior for $\phi \geq \phi_g$, and then onset of a higher modulus for $\phi \geq \phi_{RCP}$ \cite{mason95emul}.  The discrepancy in $\phi_c$ was directly addressed in simulations \cite{ikeda12}, which showed that rather than $\phi_c$ being a value that changed smoothly from 0.58 to 0.63 depending on conditions, the colloidal glass transition at $\phi_g$ and the jamming transition at $\phi_{RCP}$ are distinct transitions with different physical origins.  The rationale is that systems with larger particles are athermal, and thus have a jamming transition, whereas systems with smaller particles are thermal and see a glass transition.  For samples of intermediate particle size, two transitions may be possible.  A prior study predicted that such phenomena can be observed by measuring rheological performances of dense emulsion samples from 1 - 10 $\mu m$.\cite{ikeda12}  Indeed, the 1995 emulsion results\cite{mason95emul} support this prediction qualitatively, using droplets with mean diameter $d=1.00$~$\mu$m.  Overall, the thinking is that a thermal system will have a glass transition at the lower volume fraction $\phi_g$, but that the particles do not need to touch until they reach the higher volume fraction $\phi_{RCP}$.  For $\phi>\phi_{RCP}$, particles must deform, which is straightforward for emulsion droplets, so the rheological behavior for these large volume fractions must be dominated by the physics of the particle deformation (for example, surface tension effects for deformed emulsion droplets).  In contrast, for $\phi<\phi_{RCP}$ the rheology is determined by the thermally driven glass transition, for suitably thermal particles or droplets.

A series of experiments have explored the possibility of two distinct transitions, using  emulsions\cite{scheffold13,dinkgreve13,dinkgreve15,dinkgreve18} and colloidal systems \cite{basu14,mu15,nordstrom10,pusey04}. Most experiments focus on rheological measurements since the yield stress can be easily obtained from a plot of stress as a function of strain \cite{mewis12}. Work done primarily at the University of Pennsylvania studied thermosensitive PNIPAM colloidal particles \cite{basu14,nordstrom10}.  These samples allow for the volume fraction to be adjusted by changing the temperature.  In the second of these two papers (Basu {\it et al.} \cite{basu14}), they measured the rheological behavior of several samples and compared with their earlier results (Nordstrom {\it et al.}  \cite{nordstrom10}).  Between the two papers, the particle diameters ranged from 0.4 to 1.4 $\mu$m, to potentially cover both thermal and athermal sizes; however, the large particles used in the earlier study were softer than the smaller particles used in the later study.\cite{basu14}  The small particle samples exhibited a glass transition with $\phi_c = 0.61 \pm 0.02$, while the large particle samples had a jamming transition with $\phi_J = 0.635 \pm 0.003$.  For these samples, because the volume fraction and particle size are changed at the same time (controlled by temperature), the influence of thermal fluctuations changes by nearly a factor of two from smallest to largest volume fractions studied. Their experiments also used somewhat large steps in volume fraction ($\approx 0.01$ in Ref.~\cite{nordstrom10}, $\approx 0.05$ in Ref.~\cite{basu14}), making it challenging to precisely identify the transition points in the latter work.  

In the recent work of Dinkgreve {\it et al.},\cite{dinkgreve18} they study rheological behaviors of athermal emulsions (diameter 3.2~$\mu$m) and compare with earlier published emulsion data\cite{mason96} (diameters 0.5-1.5~$\mu$m) and earlier colloid data (diameter 0.37~$\mu$m).  The athermal samples had a yield stress for $\phi > \phi_J \approx 0.64$, and the smaller particle samples all had yield stresses for $\phi > \phi_g \approx 0.58$.  Nonetheless, they found all samples had similar scaling of their rheological curves, independent of where their transition to a yield stress was found.  However, the large and small emulsion samples used different oils (thus with different surface tensions), so it was difficult to directly compare the rheological data between the samples.

In this work we measure rheological behavior for both monodisperse and bidisperse dense emulsions with droplet diameters ranging from 1-2 $\mu m$, using same oil for the droplets, same continuous phase fluid, and same surfactant for all samples. We create our emulsions by using a seed-growth technique \cite{casper17}.  We use a weighing method to measure our samples' volume fractions, enabling consistent comparison between samples. We observe that a yield stress appears above $\phi_g=0.58$ for our samples with smaller droplet diameters ($d \approx 1$~$\mu$m) and at $\phi_J=0.63$ for our sample with the largest droplet diameter ($d \approx 2$~$\mu$m).  Using emulsions with identical compositions, and changing the volume fraction while maintaining a constant droplet size, enables us to directly compare samples with identical properties apart from volume fractions; and to compare results of droplets with different diameters but otherwise identical composition.  Our results show that indeed two distinct transitions can be seen.  Furthermore, we find that a bidisperse sample composed of both small and large droplets has comparable rheological behavior to samples composed only of small droplets.

\section{Experimental Details}
\subsection{Samples}

\begin{figure}[ht]
\centering
  \includegraphics[height=3cm]{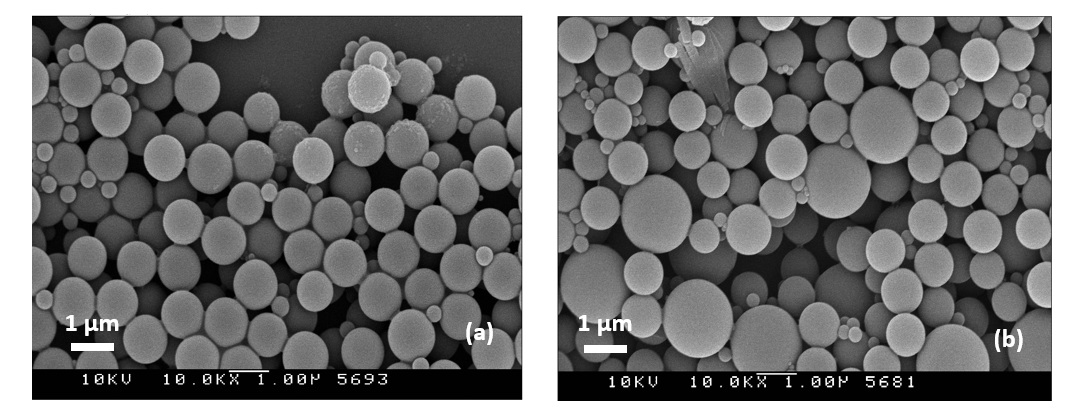}
  \caption{TPM emulsions polymerized by AIBN and observed under SEM (a) A monodisperse sample with $d_{\rm mean}$=1.16 $\mu m$. (b) Bidisperse sample with diameters $d_{\rm small}$=1.06 $\mu m$ and $d_{\rm large}$=1.86 $\mu m$. In both images, a few unusually small particles are seen (diameters under 0.2~$\mu$m), which were only observed after polymerization and which may be due to unreacted TPM oil;\cite{casper17} these small particles are not seen using optical microscopy of the emulsion samples pre-polymerization.}
  \label{sem}
\end{figure}

To prepare our 3-(Trimethoxysilyl)propyl methacrylate(TPM) samples, we use a seeded-growth method \cite{casper17} to obtain TPM emulsions at required size.  First we add 1~ml (2~ml for larger emulsions) TPM oil (3-(Trimethoxysilyl)propyl methacrylate, 98\% - Sigma-Aldrich) into 100 ml pre-made ammonia solution (1~ml 2.8\% ammonia diluted with distilled water) in a sealed plastic beaker with a stir bar.  We stir the solution at a high speed (350 rpm) to hydrolyze the oil for 20 minutes, then lower the stirring to 200 rpm to condense the oil monomers. After that, every hour we add an additional 1~ml of TPM oil until the droplets grow to our desired size. By tuning the ammonia solution's pH and the amount of added TPM oil we produce 10 ml quantities of emulsions with fairly low polydispersity.  (Note that the pH has to be measured prior to injecting the TPM oil.\cite{casper17})  We make samples with specific droplet diameters, ranging from 0.80 to 2.1 $\mu$m; see Fig.~\ref{sem}.  We then add 0.5 wt\% F108 (Synperonic F108 from Sigma-Aldrich) and 5 mM sodium chloride to stabilize our samples.

To accurately measure the sizes of our particles, we polymerize a small portion of each sample by adding 90 $\mu L$ of the emulsion solution into preheated 100~mL 0.1~wt\% $2,2$-Azobis(2-methylpropionitrile) (AIBN) solution.  We leave this mixture in a 80$^{\circ}$C oven for at least 2 hours\cite{casper17}, resulting in solidified particles due to polymerization of the TPM oil.  We then image these solid particles with a scanning electronic microscope (Topcon DS-150F Field Emission SEM); see Fig.~\ref{sem}.  We use these images to determine the diameters and polydispersity of our samples.  Previous studies suggested that the particles' volumes will shrink by about $7\%$ after polymerization \cite{casper17}, corresponding to a 2\% reduction in diameter.  To confirm that the emulsion droplet diameters are comparable to what we measure with SEM, we use differential dynamic microscopy (DDM) to measure the emulsion droplets' diameters indirectly \cite{bayles16}.  These diameters are within $5\%$ of the SEM results; we will report all diameters based on the SEM data, which has smaller uncertainty even considering the potential 2\% systematic error, that is, it is likely that the emulsion droplets are 2\% larger than the numbers we report \cite{casper17}.  Our droplets are slightly polydisperse: 6-8 $\%$; we do not ever observe our droplets to organize into crystalline arrays.

From the initial emulsions, we produce a concentrated stock emulsion by centrifuging our initial samples several times. A series of samples with lower volume fractions are then made by diluting portions of the initial samples with 0.5 wt\% F108 and 5~mM sodium chloride in water. To accurately measure the volume fractions, each sample is weighted before and after the evaporation of water to determine $\phi_w$.\cite{poon12}  As in prior work~\cite{mason96}, we correct our volume fractions to take into account the thickness $h = 17.5$~nm of the water film between two droplets pressed together near $\phi_c$, and a linear interpolation to $h=5.0$~nm at $\phi=1$.  This adjusts the volume fraction as $\phi \approx \phi_w(1+3h/2a)$ using the measured $\phi_w$ and droplet radius $a$; these are the volume fractions reported subsequently in this manuscript.  Note that we use Ref.~\cite{mason96}'s values for $h$; we do not have an independent measurement of $h$.  Due to the accuracy of our weighing, our volume fractions are correct relative to each other by $\pm 0.003$, although they may have an additional systematic uncertainty of $\pm 0.02$.\cite{poon12}

\subsection{Rheology}

Our rheological experiments are conducted using an Anton Paar MC302 rheometer. We study three monodisperse emulsions ($d_{\rm mean}$= 1.03, 1.16, and 2.03 $\mu m$) and one bidisperse sample ($d_{\rm small}$=1.06 $\mu m$, $d_{\rm large}$=1.86 $\mu m$, with 1:1 ratio in volume) at room temperature (25$^{\circ}$C ). For each sample, we perform a steady shear measurement with a 50 mm cone-plate geometry (truncation height 53 $\mu m$, cone angle 1.01$^{\circ}$). A solvent trap is used to minimize sample evaporation. A 50~mm diameter rough bottom plate is used to eliminate slip at the plate; the cone is not roughened. To provide a reproducible initial condition, we pre-shear all samples with a 10 s$^{-1}$ shear rate for 30~s and then sit still for another 30~s. All measurements are performed under room temperature. Sedimentation and creaming of particles is negligible within our experimental time scale.

\section{Results}
\label{sectionResults}

\begin{figure}[ht]
 \centering
 \includegraphics[width=10cm,keepaspectratio]{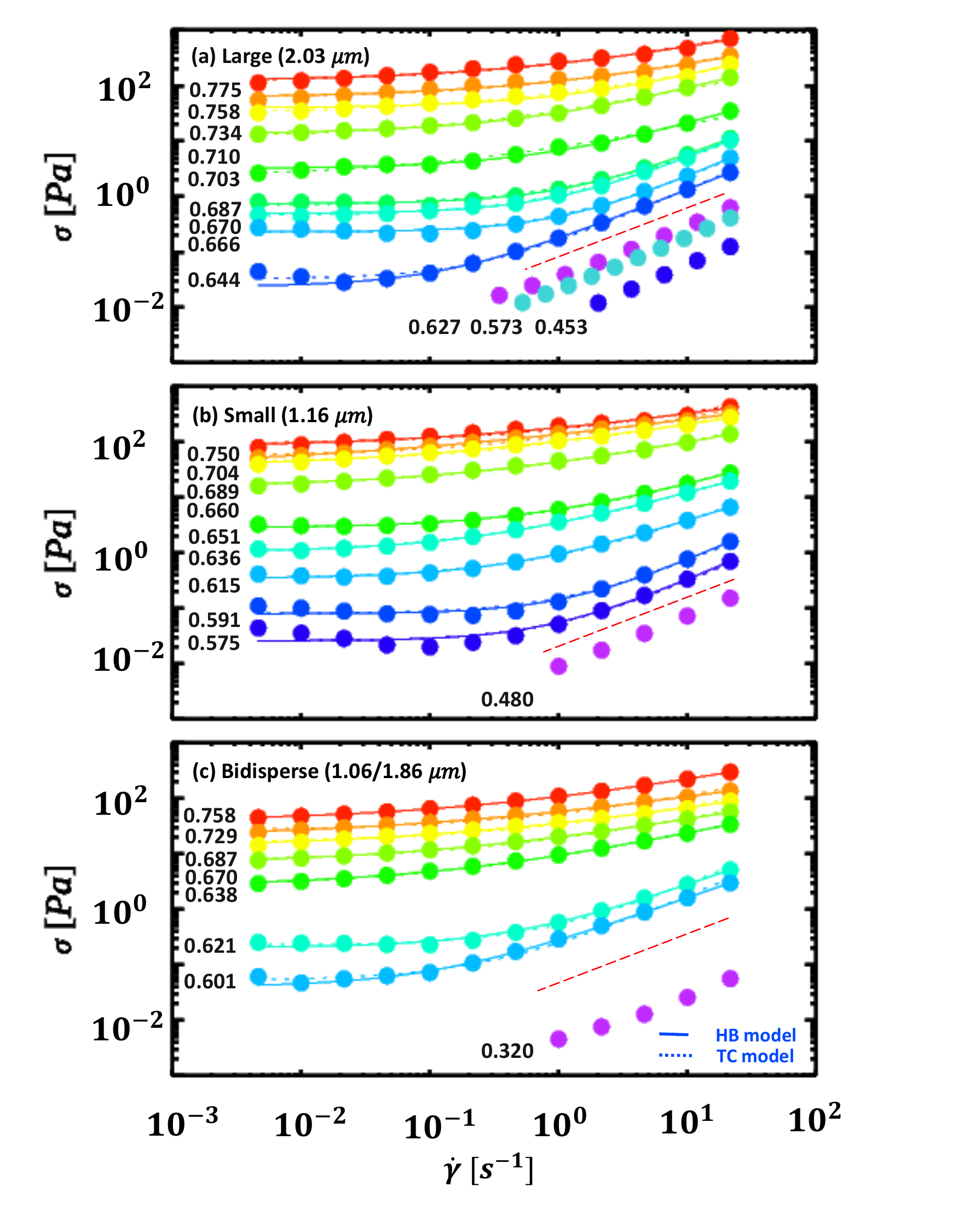}
 \caption{Shear stress $\sigma$ plotted as a function of strain rate $\dot{\gamma}$ for (a) the monodisperse sample with droplet diameter $d=2.03$~$\mu$m, (b) the monodisperse sample with $d=1.16$~$\mu$m, (c) the bidisperse sample with $d_{\rm small}=1.06$~$\mu$m, $d_{\rm large}=1.86$~$\mu$m.  The curves are labeled by their volume fractions $\phi$.  The solid lines are fitting curves with the HB model (Eq.~\ref{hb}), and the dotted lines are fitting with the TC model (Eq.~\ref{tc}). \cite{marco20}  The red dashed lines are guidelines that identify the transition between samples with and without a yield stress.}
 \label{rheo}
\end{figure}

Figure \ref{rheo} shows the rheological curves (stress as a function of strain rate) for two monodisperse samples (panels a and b) and the bidisperse sample (panel c). For all three samples, steady shear measurements are performed with shear rates ranging from $\dot{\gamma} = 10^2$ to 10$^{-3}$~s$^{-1}$.  For high volume fractions $\phi > \phi_c$ we observe a yield stress, signaled by a finite value of stress $\sigma$ as $\dot{\gamma} \rightarrow 0$.  The existence of this yield stress is consistent with previous experiments \cite{mason95emul,dinkgreve18,bonn17}. 

The curves of Fig.~\ref{rheo}(a,b) show that the transition from solid to liquid happens at different $\phi$ values for these two different monodisperse samples.  For our large droplet sample ($d_{\rm mean}=2.03$~$\mu$m), $\phi=0.643$ has a yield stress and $\phi=0.627$ does not, indicating that the transition volume fraction $\phi_c$ lies between these two values.  This $\phi_c = 0.635 \pm 0.008$ is similar to results observed in granular systems \cite{brodu15}.  For our smaller droplet sample ($d_{\rm mean} = 1.16$~$\mu$m), the data indicate $\phi_c < 0.575 \approx \phi_g$, suggesting that this sample is glassy for $\phi \geq 0.575$.  Prior work argued that it is hard to accurately compare volume fractions measured by employing different methods \cite{poon12}, and as mentioned above all of our volume fractions likely have a systematic uncertainty of $\pm 0.02$; however, the key point is that we use a consistent weight-measuring method for all of our samples, and the relative uncertainties of our stated volume fraction values is $\pm 0.003$.  In some prior work, due to differing volume fraction measurement techniques, the uncertainty for $\phi_c$ between different groups was as much as $\pm 0.03-0.06$, raising the possibility that one group could measure $\phi_c=0.58$ and another group could measure $\phi_c=0.63$ and these could potentially be equivalent \cite{poon12}.  Here, since we use a consistent method for determining $\phi$ across all of our samples, we have strong evidence that the solid-to-liquid transition occurs at lower volume fraction for the smaller particles. This supports the idea that smaller droplets have more significant thermal fluctuations, resulting in a glass transition at a lower volume fraction, whereas the larger droplets are more athermal \cite{ikeda12}.

\begin{figure*}
 \centering
 \includegraphics[width=18cm,keepaspectratio]{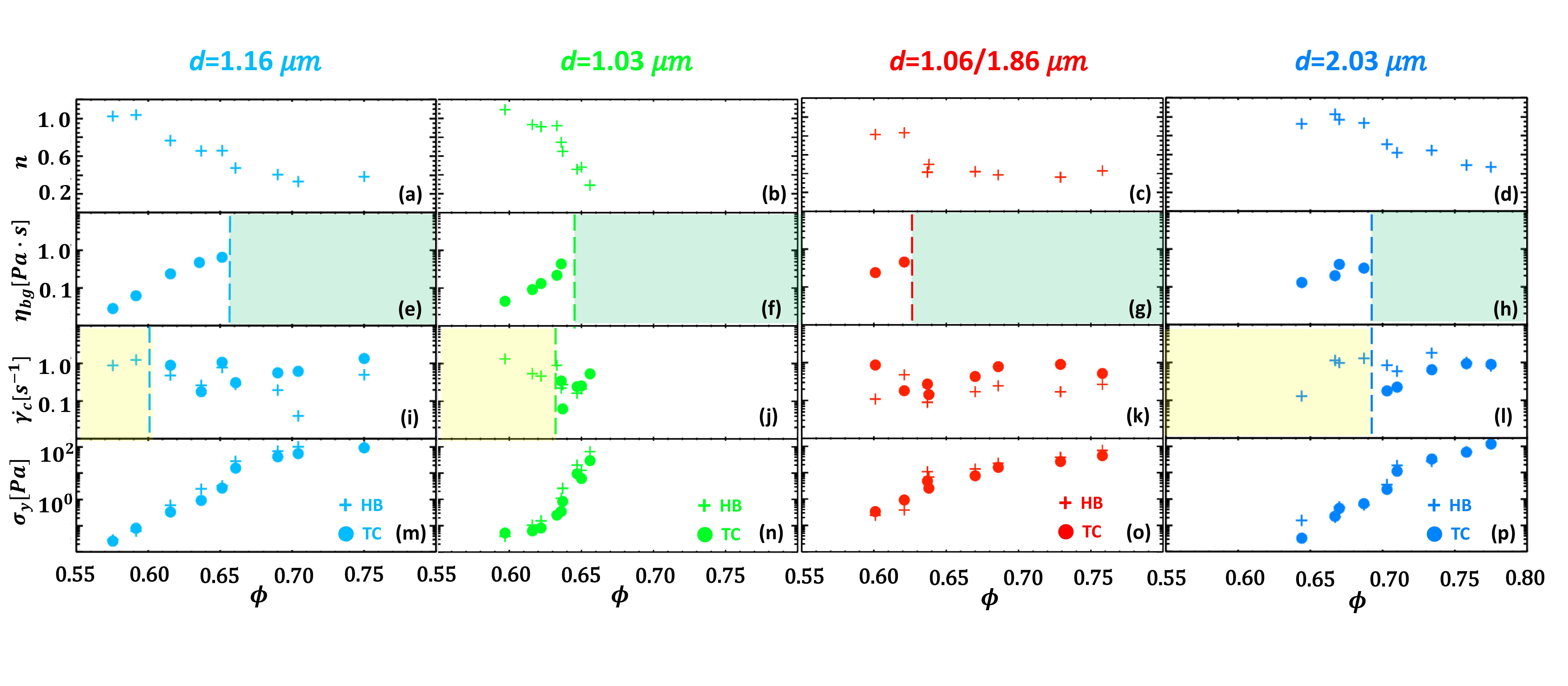}
 \caption{The Herschel–Bulkley (HB) and three component (TC) model fitting parameters as a function of volume fraction $\phi$ for monodisperse samples with small droplets (two left columns), a bidisperse sample (third column), and large droplets ($d=2.03$~$\mu$m, right column).  (a-d) The HB $n$ parameter.  (e-h) The TC background viscosity parameter $\eta_{bg}$.  The green shaded regions indicate cases for which setting $\eta_{bg}=0$ results in a less than 1\% increase in the least squares fitting error.  (i-l) The TC and HB characteristic strain rate parameter $\dot{\gamma_{\rm c}}$.  The yellow shaded regions indicate cases for which setting $\dot{\gamma_{\rm c}} \rightarrow \infty$ results in a less than 1\% increase in the least squares fitting error.  (m-p) The yield stress $\sigma_{\rm y}$ from the two models.
 }
 \label{rheofittc} 
\end{figure*}

All the stress-strain data for $\phi > \phi_{\rm c}$ is well fit with the Herschel–Bulkley (HB) model: 
\begin{equation}
\sigma=\sigma_{\rm y}+k\dot{\gamma}^n.
\end{equation}
\noindent These fits are the solid lines in Fig.~\ref{rheo}.  $\sigma_y$ is the yield stress, $k$ is consistency index, and $n$ is the flow index.  The flow index $n$ is plotted as a function of volume fraction $\phi$ in Fig.~\ref{rheofittc}(a-d).  For all samples, $n$ is close to 1 for the lower volume fractions ($\phi \approx \phi_c$) and then decreases for larger volume fractions, reaching around $n \approx 0.4-0.5$ for the largest volume fractions we consider.  We also show the yield stress $\sigma_y$ as the $+$ symbols in Fig.~\ref{rheofittc}(m-p).  As above, the samples composed of smaller droplets shows a nonzero $\sigma_y$ starting at $\phi \approx 0.57$ [Fig.~\ref{rheofittc}(m,n)] while the sample composed of larger droplets shows a nonzero $\sigma_y$ starting at $\phi \approx 0.64$ [Fig.~\ref{rheofittc}(p)].  Rather than considering the consistency index $k$ which has unusual units, and anticipating a different fitting model, we rewrite the HB equation as
\begin{equation}
\sigma=\sigma_{\rm y}+\sigma_{\rm y}(\dot{\gamma}/\dot{\gamma_{\rm c}})^{n}.
\label{hb}
\end{equation}
Here we have replaced the fitting parameter $k$ with $\dot{\gamma_{\rm c}}$, which is a characteristic strain rate scale that will be discussed below.

The different fitting model we next use to describe our data is the ``three-component'' (``TC'') model\cite{marco20}, which has the form:
\begin{equation}
\sigma=\sigma_{\rm y}+\sigma_{\rm y}(\dot{\gamma}/\dot{\gamma_{\rm c}})^{1/2}+\eta_{\rm bg}\dot{\gamma}.
\label{tc}
\end{equation}
These fits are indicated by the dashed lines in Fig.~\ref{rheo}.  Again, $\sigma_{\rm y}$ is the yield stress, representing an elastic component for $\dot{\gamma} \rightarrow 0$; the circles in Fig.~\ref{rheo}(m-p) show that this yield stress is quite similar to the HB fit data ($+$ symbols in those panels). The second component is plasticity, with fit parameter $\dot{\gamma_{\rm c}}$ which is the strain rate at which the contributions to the stress from plasticity and elasticity are equal.  The third component is viscous behavior, with $\eta_{\rm bg}$ being the background viscosity -- presumably the viscosity of the continuous phase in our emulsion samples, but in practice a third fitting parameter.  Figure \ref{rheo} shows that for both samples, the TC model and HB model fit our original data nearly equally well, with less than $10\%$ difference in the least squares fitting error.

In addition to the yield stress, the remaining fitting parameters of the TC model are shown in Fig.~\ref{rheofittc}(e-l). For the monodisperse samples with small droplets [Fig.~\ref{rheofittc}(e,f)] we notice that above the jamming transition ($\phi>0.64$), $\eta_{\rm bg}$ decreases dramatically, showing that viscous behavior plays little role in the high volume fraction samples, at least for the strain rates we can measure.  For situations where the error bar of $\eta_{\rm bg}$ does not exclude zero, we do not plot a point.  The abrupt decrease in $\eta_{\rm bg}$ for $\phi>0.65$ may be related to a jamming transition within the sample.  This simplification of Eq.~\ref{tc}, with just the first two components, matches the prediction of the kinetic elastoplastic model, which appropriately is expected to apply for jammed materials.\cite{bocquet09,mansard13}  This rheological behavior has often been seen in high volume fraction foams; see Ref.~\cite{denkov09} for a review.  It corresponds to the HB model with $n=1/2$.  Note that the values of $\eta_{\rm bg}$ we find from the fitting are significantly larger than the true background (continuous) phase viscosity ($\eta = 1.5 \times 10^{-3}$ Pa$\cdot$s), and likewise larger than the viscosity of the oil in the droplets ($\eta = 2.0\times 10^{-3}$ Pa$\cdot$s).

Similar to the transition in $\eta_{\rm bg}$, we observe an abrupt transition when plotting the TC fit parameter $\dot{\gamma_{\rm c}}$ as a function of $\phi$ in Fig.~\ref{rheofittc}(i,j): $\dot{\gamma_{\rm c}}$ grows dramatically for $\phi<0.61$, essentially setting the contribution of the plastic term to zero.  Again, where the uncertainty of $\dot{\gamma_{\rm c}}$ does not exclude $\infty$, we do not plot a point.  In this case the TC model (Eq.~\ref{tc}) reduces to the Bingham model.  For the HB model, $\dot{\gamma_{\rm c}}$ is always finite.  For the two samples with small droplets, Fig.~\ref{rheofittc}(e,f,i,j) show that there is a range of $\phi$ for which  both plastic and viscous behavior coexist; both terms are needed to describe the stress/strain data.

For the sample with larger droplets [right-most column of Fig.~\ref{rheofittc}], we also observe those abrupt changes in $\dot{\gamma_{\rm c}}$ and $\eta_{\rm bg}$. Interestingly, both of these changes happen at around same volume fraction $\phi \approx 0.69$, suggesting there is no volume fraction at which both plastic and viscous behaviors are observable in this athermal sample. By comparing the HB and TC fitting parameters, we notice that both models suggest plastic behavior dominates at high volume fractions ($n \approx 0.5$ and $\eta_{\rm bg}=0$) and viscous behavior plays a more important role for samples close to the glass transition ($n \approx 1$ and $\dot{\gamma_{\rm c}} \rightarrow \infty$).

Finally, for the bidisperse sample (red data, third column of Fig.~\ref{rheofittc}) we see behavior mostly similar to the two samples with smaller droplets.  The similarities include that $\eta_{\rm bg}$ vanishes at $\phi \approx 0.63$, and $\sigma_y$ is nonzero for $\phi \gtrsim 0.59$.  Intriguingly, we do not see any range for which the TC fit parameter $\dot{\gamma_{\rm c}}$ can be neglected, in contrast with the other three samples.

\begin{figure}
\centering
\includegraphics[width=9cm,keepaspectratio]{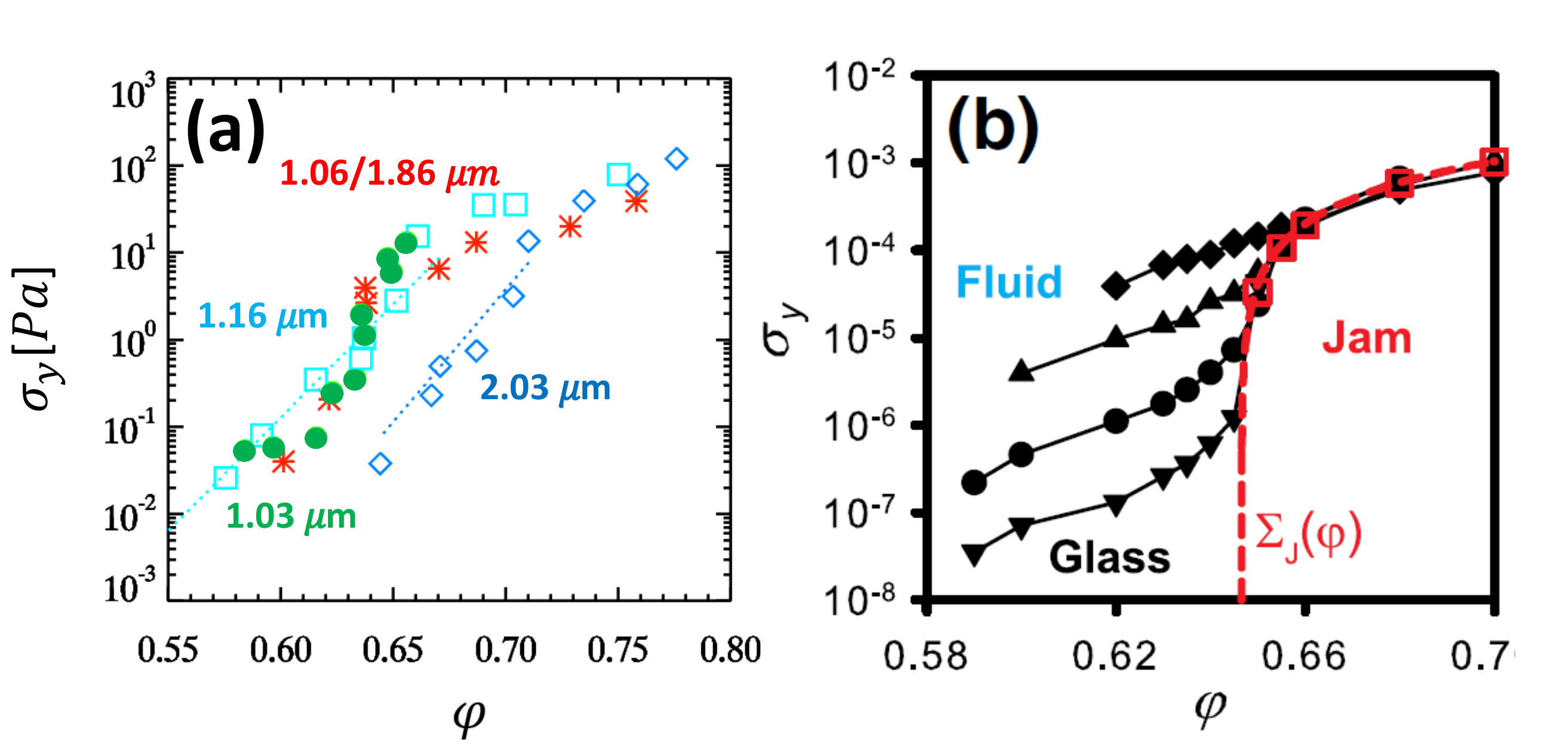}
\caption{(a) The yield stress $\sigma_{\rm y}$ as a function of volume fraction $\phi$ for our experimental data; each set of data is labeled by the mean droplet diameter.  The red asterisk data are for the bidisperse sample, labeled by the two mean droplet diameters.  The effective temperature ranges from $T_{\rm eff} = 5\times 10^{-6}$ for the $d=1.03$~$\mu$m sample to $1.3\times 10^{-6}$ for the $d=2.03$~$\mu$m sample; see text for details.  The dashed lines are exponential fit curves $\sigma_y = \sigma_0 e^{C \phi}$ with $\sigma_0=3 \times 10^{-17}$ Pa, $C=60$ for small particles and $\sigma_0=2 \times 10^{-21}$ Pa, $C=70$ for the larger particles.\cite{ken05}  (b) Simulation results from Ref.~\cite{ikeda12}.  As $\phi$ is decreased, the system goes through the jamming transition at around 0.64, with an abrupt drop in yield stress. After that, it will stay in a glassy regime, with only a moderate decrease of the yield stress until it reaches the glass transition point $\phi_g \approx 0.58$, at which point the yield stress vanishes \cite{ikeda12}. The red dashed line represents the $T=0$ limit (for large, athermal particles).  For the black curves, the effective temperature varies by factors of ten, ranging from the top black curve (``Fluid,'' $T_{\rm eff} = 10^{-4}$ to the bottom black curve (``Glass,'' $T_{\rm eff} = 10^{-7}$).
}
\label{sigmavsphi} 
\end{figure}

To compare with the simulation predictions from Ikeda {\it et al.},\cite{ikeda12} we plot $\sigma_{\rm y}$ as a function of $\phi$ for all our samples in Fig.~\ref{sigmavsphi}(a).  As we observe no distinct differences between the yield stresses obtained by HB and TC model, for simplicity, we use the HB $\sigma_{\rm y}$.  In our experiment, for the small droplet diameter samples (1.03, 1.16 $\mu m$), the yield stress only weakly depends on $\phi$ for high $\phi$.  A more significant change in $\sigma_{\rm y}$ is seen around $\phi \approx 0.65$, although $\sigma_{\rm y}$ still remains nonzero until $\phi \lesssim 0.58$.  This qualitatively resembles Ikeda {\it et al.}'s ``glass'' simulation data shown in  Fig.~\ref{sigmavsphi}(b).  In contrast, our large diameter droplet sample (2.03 $\mu m$) has a yield stress which starts to decrease rapidly at higher volume fraction (at around $\phi \approx 0.70$), and $\sigma_{\rm y}$ disappears at $\phi < 0.643$. These results strongly suggest that the critical solid-to-liquid transition happens at different volume fractions depending on particles' size.  For the simulations, the $\sigma_{\rm y}$ data overlap for $\phi \gtrsim 0.66$ whereas for our experiment, overlap is only see for $\phi \gtrsim 0.70$.

To characterize the difference between ``large athermal'' and ``small thermal'' particles, Ikeda {\it et al.} considered the effective temperature $T_{\rm eff} = k_B T/\varepsilon$, where $\varepsilon$ represents an energy scale related to the particle stiffness.\cite{ikeda12}  This effective temperature characterizes how easy it is for thermal fluctuations to deform particles, thus allowing them to slip past one another. In our experiment, we measure the TPM surface tension using the pendant drop method (Dropometer, made by Dropletlab).  We measure the surface tension to be $\Sigma=3$~mN/m, which is consistent with previous work.\cite{kraft11}  We assume the deformation energy $\varepsilon = \Sigma d^2$, which should be the correct order of magnitude.  Using this we get $k_B T / \varepsilon = (5.2 - 20.0) \times 10^{-6}$, lying in the crossover regime predicted by the simulation. \cite{ikeda12}  The main qualitative difference between our results [Fig.~\ref{sigmavsphi}(a)] and the simulation results [Fig.~\ref{sigmavsphi}(b)] is that our large droplet sample depends more strongly on volume fraction for $\phi>\phi_J$:  a fairly smooth decrease in $\phi_J$ by several orders of magnitude is seen as $\phi$ decreases from 0.77 to 0.65.

A more cautious approach suggests that the effective temperature may be larger than the above considerations.  The total surface energy of a droplet is $\Sigma \pi d^2$, but thermal fluctuations need not create these droplets from nothing; rather, the concept is that the particles can move past one another due to thermal fluctuations resulting in slight size changes.  Conserving the volume of a droplet, a decrease in diameter $\Delta d$ along one axis results in an increase along the other axes of $\Delta d/2$ (assuming the diameter fluctuations are small, $\Delta d \ll d$).  This results in a change of droplet surface area by $\Delta A \sim (\Delta d)^2$.  If a diameter fluctuation $\Delta d/d = 0.1$ is sufficient to allow a droplet to move past another -- and thus for the sample to flow -- then the necessary surface energy change is of order $\varepsilon = 0.01 \pi \Sigma d^2$, a factor of 30 smaller than our earlier estimate of $\varepsilon = \Sigma d^2$.  This suggests our effective temperatures may be in the range $k_B T/ \varepsilon = (1.7 - 6.4) \times 10^{-4}$.

Intriguingly, as noted above, the yield stress of our bidisperse sample with droplet diameters 1.06~$\mu$m and 1.86~$\mu$m behave similarly to the two small droplet size monodisperse samples [Fig.~\ref{sigmavsphi}(a)].   This suggests that in a bidisperse sample, the small droplets dominate the rheological behavior.  This seems sensible:  the glass transition should be induced by the thermal motion of the small particles.  In our bidisperse sample, the two droplets have equal volume fractions within the sample; it is an open question as to how much of the smaller droplet is necessary to see a glass transition.

For samples with $\phi > \phi_c$, we fit the yield stress data to an exponential growth model; see the straight lines in Fig.~\ref{sigmavsphi}(a).  These fits are just to the data where the growth of $\sigma_y$ appears roughly linear on this semilog plot, so $0.58 \leq \phi \leq 0.64$ for the small diameter emulsion sample and $0.64 \leq \phi \leq 0.70$ for the large diameter emulsion sample.   In our experiments, the yield stress grows with volume fraction more strongly when increasing particle's diameter, from $\sigma _y \sim C_1 e^{60\phi}$ to $\sigma _y \sim C_2 e^{70\phi}$. These fits are consistent with the entropic barrier hopping model suggested by Kobelev and Schweizer \cite{ken05}, despite slight differences in the exponent ($\sigma_y \sim e^{40\phi}$ for them). Consistent with their prediction, the larger the particles are, the more abruptly $\sigma_y$ will decrease with $\phi$.  The prefactors have values $C_1 = 3 \times 10^{-17}$ and $C_2 = 2 \times 10^{-21}$, and are not assigned any physical meaning.\cite{ken05}

\begin{figure}
\centering
\includegraphics[width=8cm,keepaspectratio]{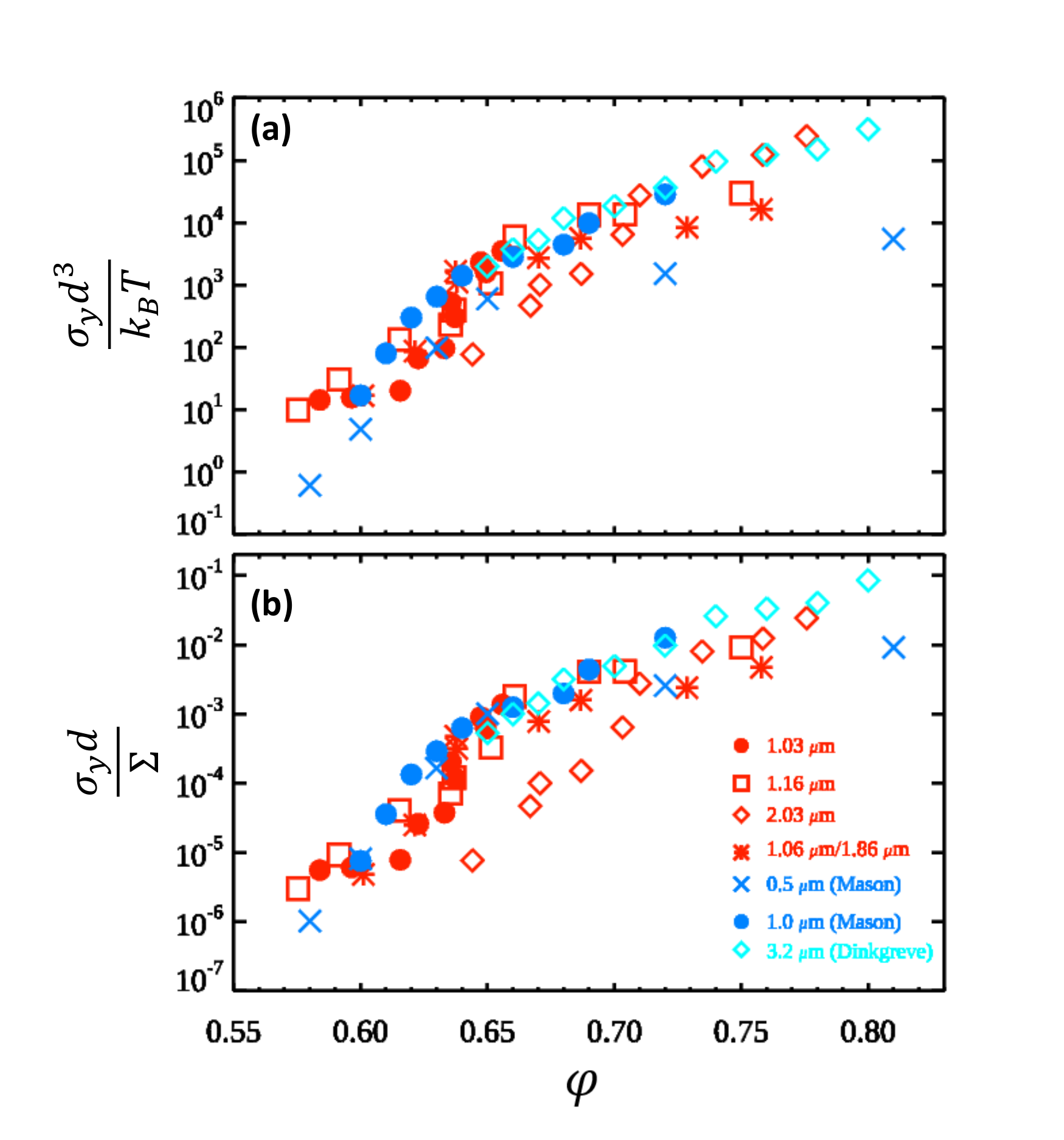}
\caption{(a) The yield stress as a function of $\phi$, with yield stress nondimensionalized by the thermal energy $k_B T$.  (b) The yield stress as a function of $\phi$, with yield stress nondimensionalized by the oil-water surface tension $\Sigma$.  The legend indicates the droplet diameter $d$ and the source of the data, if not our work.\cite{mason96,dinkgreve18}  For our bidisperse sample, we use $d=1.06$~$\mu$m to scale the data, consistent with this sample appearing rheologically similar to our other two samples with $d \approx 1$~$\mu$m.}
\label{nonyield} 
\end{figure}
%(a) thermal yield stress VS volume fraction $\phi$  (b) mechanical yield stress VS $\phi$

To further understand how the yield stress changes with volume fraction, we consider three additional data sets from prior literature.  Two data sets are taken from Mason, Bibette, and Weitz (Ref.~\cite{mason96}) who studied silicon oil in water emulsions using both steady and oscillatory shear.  These emulsions were stabilized by sodium dodecyl sulfate (SDS) and employed the fractionation method to produce fairly monodisperse samples with droplet diameters between 0.5~$\mu$m to 1.5~$\mu$m; the surface tension was $\Sigma=9.8$~mN/m.  They investigated the yield stress of their samples.  They concluded that their samples had a size-independent transition which they called a glass transition at $\phi = 0.58$.  The third data set is from Dinkgreve {\it et al.}\cite{dinkgreve13,dinkgreve15,dinkgreve18} who studied castor oil in water emulsions using steady shear, similar to our experiments.  Their samples had a larger mean droplet diameter, $d=3.2$~$\mu$m, and the surface tension was 1.5 mN/m; the droplet size was intentionally chosen to be in the athermal regime.  The large athermal droplets had a liquid-to-solid jamming transition at $\phi = 0.64$.  Intriguingly, in Ref.~\cite{dinkgreve18} they also examined the earlier data of Mason, Bibette, and Weitz,\cite{mason96} and found that all the data (both thermal and athermal droplet sizes) fit well into an identical master curve for $\phi > \phi_c$, albeit with different $\phi_c$ values for the different droplet sizes.

% this is the "mobile" data from Dinkgreve.

We compare these prior data sets to our emulsion data Fig.~\ref{nonyield}.  To better understand how thermal motion and particle deformation affect our system, here we define a nondimensional thermal yield stress $\sigma_{\rm T}=\sigma_{\rm y}d^3/k_{B}T$ and a nondimensional mechanical yield stress $\sigma_{\rm 0}=\sigma_{\rm y}d/\Sigma$.  Our data are the red symbols, and the literature data are the blue symbols, as indicated by the legend in Fig.~\ref{nonyield}(b).  In all cases, the effective temperature is small ($k_{B}T/\varepsilon \ll 1$).  Neither the thermal yield stress [Fig.~\ref{nonyield}(a)] nor the mechanical yield stress [Fig~\ref{nonyield}(b)] collapse the data perfectly.

The thermal yield stress comparison is based on a simple idea, that the modulus of a solid should scale as $k_B T / d^3$ for a sample made of components with inter-particle distance $d$ \cite{jones02} -- indeed, this is the basic reason that ``soft matter'' composed of micron-sized objects is softer than ``hard matter'' made of atoms with nanometer spacing.  This also should be the relevant yield stress for the glass transition, as at $\phi = \phi_g \approx 0.58$, the droplets do not need to touch each other and certainly do not need to deform.  Thus, the surface tension should be irrelevant at this volume fraction, and the influence of thermal motion should dominate the yield stress, which is essentially why the glass transition is regarded as only present for thermal particles.  The thermal yield stress somewhat collapses the data, with the outliers being the smallest diameter droplets ($d=0.5$~$\mu$m data from Ref.~\cite{mason96}).  However, even with our own data, the thermal yield stress varies by an order of magnitude for constant $\phi$.  Intriguingly, the largest droplet samples (the red diamonds with $d=2.03$~$\mu$m and the blue diamonds with $d=3.2$~$\mu$m) agree reasonably well with the other data for $\phi > 0.64$, that is, the volume fractions where these large droplet samples have a yield stress -- but this is not the range where we would expect the best collapse for the thermal yield stress.

Close to $\phi_g$, we have $\sigma_{\rm y}d^3/k_B T = O(10^1)$ for our samples [red symbols in Fig.~\ref{nonyield}(a)].  Earlier experiments with hydrogel particles found this to be $O(10^{0})$ for small ($d \approx 0.5$~$\mu$m) softer particles and $O(10^2)$ for larger ($d\approx 1.5$~$\mu$m) harder particles.\cite{basu14,nordstrom10}.  Given the variability of the earlier results, it is not surprising we are of similar order of magnitude.

In Fig.~\ref{nonyield}(b) we plot the mechanical yield stress; the idea is that the yield stress is due to the surface tension for $\phi>\phi_{RCP}=0.64$ as at these volume fractions, the droplets be deformed.  The surface energy of that deformation gives rise to the macroscopic elastic response.  At or below the yield stress these slight deformations are sufficient to allow droplets to move past one another and the sample can flow.  Essentially, using the mechanical stress is equivalent to the simulation prediction from Fig.~\ref{sigmavsphi}(b) where the data from all five curves collapse for $\phi \gtrsim 0.66$ -- above this point, for our experiment, all yield stresses should be set by the mechanical yield stress.  Our samples indeed collapse fairly well for $\phi \gtrsim 0.72$ [red symbols in Fig.~\ref{nonyield}(b)].  All of the data (ours and the literature data) collapse fairly well even for lower volume fractions, with the sole exception being our large droplet sample with $d=2.03$~$\mu$m.  The agreement of our samples at high volume fractions is sensible as the droplets are moderately deformed in all cases, and thus a yield stress more strongly determined by the surface tension.\cite{princen83}  It is possible that the disagreement at lower volume fractions is another sign that our large droplet sample has a jamming transition rather than a glass transition, but in that case it is surprising that the still larger $d=3.2$~$\mu$m sample from Dinkgreve {\it et al.} collapses with the smaller droplet data rather than our large droplet data.  To summarize Fig.~\ref{nonyield}, it is worth noting that for our own data at least, $k_B T$ and $\Sigma$ are constant, so the difference between the two scalings is a factor of $d^3$ in (a) and $d^1$ in (b).

\section{Conclusions}

We have studied the rheological behavior of concentrated emulsions over a range of droplet sizes and volume fractions, finding evidence of both a thermal-like glass transition at $\phi_g \approx 0.58$ and an athermal-like jamming transition at $\phi_J \approx 0.64$.  For the samples with a glass transition, the yield stress becomes nonzero for $\phi > \phi_g$, while for samples without a glass transition, we do not measure a nonzero yield stress until $\phi > \phi_J$.  The glass transition is observable in our samples with smaller droplet diameters ($d \approx 1$~$\mu$m), whereas our sample with the largest diameter ($d = 2.03$~$\mu$m) is the case with only a jamming transition.  We additionally find evidence that our samples with a glass transition also have a jamming transition at $\phi_J$.  This is marked by a dramatic rise of the yield stress by about two orders of magnitude and a disappearance of the viscous component from the three-component (TC) model fit (Eq.~\ref{tc}).  Intriguingly, the ``athermal'' sample with only a jamming transition shows a transition of a different sort around $\phi \approx 0.70$, where the TC model fit transitions from needing only a viscous component ($\phi < 0.70$) to needing only a plastic component ($\phi > 0.70$).  Of course, it is possible that the apparently missing components simply mean that we cannot measure them with the resolution of our rheometer; in particular, the viscous component may simply have moved to shear rates faster than our ability to measure, as certainly the limiting behavior for $\dot{\gamma} \rightarrow \infty$ must be viscous.\cite{trappe00}

The key advantages of our experiments are that we use one type of particle for all measurements, emulsion droplets of the same oil in water, so that the particle interaction does not change as we change the droplet size; we have a consistent means of measuring the volume fraction across all samples; and we vary volume fraction while keeping our droplet size fixed (in contrast to hydrogel particles, for example \cite{basu14}).  A concern might be that we do not have a sample with only a glass transition, and not a jamming transition.  However, as per Fig.~\ref{sigmavsphi}(b), our samples have deformable droplets and the volume fraction can always be increased well above $\phi_J$, and thus we should always see a jamming transition around $\phi_J$ once the droplets begin to deform -- which indeed is what is suggested by our data in Fig.~\ref{sigmavsphi}(a).

Prior simulations by Olsson and Teitel found that the rheological curves for soft particles near jamming could be rescaled onto master curves (one curve for jammed samples, a second curve for unjammed samples).\cite{teitel07,olsson11}  Given that our data are well-fit with both the HB and TC models -- but with varying power law exponent $n$ (HB model) or other fitting coefficients (TC model), we have not presented a data collapse of the raw rheological data shown in Fig.~\ref{rheo}.  To the extent that the HB and TC models are reasonable fits, the dependence of the fitting parameters on $\phi$ shows that our data do not follow a master curve.  This is in contrast with some prior experimental work with hydrogel particles, for which $n\approx 0.5$ was essentially constant, allowing for a good data collapse.\cite{basu14,nordstrom10}  In contrast to our work, these soft hydrogel particles only had one transition (a glass transition) as their volume fraction was increased, even to well above $\phi_{\rm RCP}$.  The deformability of hydrogel particles is not due to surface tension; it is somewhat due to the elasticity of the particles, but with additional complications intrinsic to their polymeric nature.\cite{stjohn07,lyon12}.  Nonetheless one might expect that at $\phi_g$ the hydrogel particles would not yet be deformed and that their rheological behavior would change for $\phi > \phi_J$ after deformations are mandatory; this was not seen.\cite{basu14,nordstrom10}.  Another interesting difference is that our yield stresses increase by three to four orders of magnitude as we increase $\phi$ (see for example Fig.~\ref{sigmavsphi}), in agreement with the early work of Mason and Weitz,\cite{mason95emul} whereas the hydrogel particles saw an increase of only one order of magnitude over a similar range of $\phi$.\cite{basu14}.  Comparing with our work, these interesting differences suggest that liquid-to-solid transitions as the volume fraction is increased may be non-universal in ways beyond the size-dependent glass transition / jamming transition distinction; that the particle type matters as well.
 
\section*{Conflicts of interest}
There are no conflicts to declare.

\section*{Acknowledgements}
We thank J.C. Crocker, C. Orellana, and K.V. Edmond for helpful conversations.
The work of C.C. and E.R.W. was supported by the NSF under Grant No. DMR-1609763.

%%%END OF MAIN TEXT%%%


\balance


%%%REFERENCES%%%
\bibliography{rsc} %You need to replace "rsc" on this line with the name of your .bib file

\providecommand*{\mcitethebibliography}{\thebibliography}
\csname @ifundefined\endcsname{endmcitethebibliography}
{\let\endmcitethebibliography\endthebibliography}{}
\begin{mcitethebibliography}{38}
\providecommand*{\natexlab}[1]{#1}
\providecommand*{\mciteSetBstSublistMode}[1]{}
\providecommand*{\mciteSetBstMaxWidthForm}[2]{}
\providecommand*{\mciteBstWouldAddEndPuncttrue}
  {\def\EndOfBibitem{\unskip.}}
\providecommand*{\mciteBstWouldAddEndPunctfalse}
  {\let\EndOfBibitem\relax}
\providecommand*{\mciteSetBstMidEndSepPunct}[3]{}
\providecommand*{\mciteSetBstSublistLabelBeginEnd}[3]{}
\providecommand*{\EndOfBibitem}{}
\mciteSetBstSublistMode{f}
\mciteSetBstMaxWidthForm{subitem}
{(\emph{\alph{mcitesubitemcount}})}
\mciteSetBstSublistLabelBeginEnd{\mcitemaxwidthsubitemform\space}
{\relax}{\relax}

\bibitem[Bonn \emph{et~al.}(2017)Bonn, Denn, Berthier, Divoux, and
  Manneville]{bonn17}
D.~Bonn, M.~M. Denn, L.~Berthier, T.~Divoux and S.~Manneville, \emph{Rev. Mod.
  Phys.}, 2017, \textbf{89}, 035005\relax
\mciteBstWouldAddEndPuncttrue
\mciteSetBstMidEndSepPunct{\mcitedefaultmidpunct}
{\mcitedefaultendpunct}{\mcitedefaultseppunct}\relax
\EndOfBibitem
\bibitem[Mason \emph{et~al.}(2006)Mason, Wilking, Meleson, Chang, and
  Graves]{mason06}
T.~G. Mason, J.~N. Wilking, K.~Meleson, C.~B. Chang and S.~M. Graves,
  \emph{Journal of Physics: Condensed Matter}, 2006, \textbf{18},
  R635--R666\relax
\mciteBstWouldAddEndPuncttrue
\mciteSetBstMidEndSepPunct{\mcitedefaultmidpunct}
{\mcitedefaultendpunct}{\mcitedefaultseppunct}\relax
\EndOfBibitem
\bibitem[Liu and Nagel(1998)]{andrea98}
A.~J. Liu and S.~R. Nagel, \emph{Nature}, 1998, \textbf{396}, 21--22\relax
\mciteBstWouldAddEndPuncttrue
\mciteSetBstMidEndSepPunct{\mcitedefaultmidpunct}
{\mcitedefaultendpunct}{\mcitedefaultseppunct}\relax
\EndOfBibitem
\bibitem[O'Hern \emph{et~al.}(2003)O'Hern, Silbert, Liu, and Nagel]{ohern03}
C.~S. O'Hern, L.~E. Silbert, A.~J. Liu and S.~R. Nagel, \emph{Phys. Rev. E},
  2003, \textbf{68}, 011306\relax
\mciteBstWouldAddEndPuncttrue
\mciteSetBstMidEndSepPunct{\mcitedefaultmidpunct}
{\mcitedefaultendpunct}{\mcitedefaultseppunct}\relax
\EndOfBibitem
\bibitem[Liu and Nagel(2010)]{andrea10}
A.~J. Liu and S.~R. Nagel, \emph{Ann. Rev. Cond. Mat. Phys.}, 2010, \textbf{1},
  347--369\relax
\mciteBstWouldAddEndPuncttrue
\mciteSetBstMidEndSepPunct{\mcitedefaultmidpunct}
{\mcitedefaultendpunct}{\mcitedefaultseppunct}\relax
\EndOfBibitem
\bibitem[Desmond and Weeks(2014)]{desmond14}
K.~W. Desmond and E.~R. Weeks, \emph{Phys. Rev. E}, 2014, \textbf{90},
  022204\relax
\mciteBstWouldAddEndPuncttrue
\mciteSetBstMidEndSepPunct{\mcitedefaultmidpunct}
{\mcitedefaultendpunct}{\mcitedefaultseppunct}\relax
\EndOfBibitem
\bibitem[Pusey and van Megen(1986)]{pusey86}
P.~N. Pusey and W.~van Megen, \emph{Nature}, 1986, \textbf{320}, 340--342\relax
\mciteBstWouldAddEndPuncttrue
\mciteSetBstMidEndSepPunct{\mcitedefaultmidpunct}
{\mcitedefaultendpunct}{\mcitedefaultseppunct}\relax
\EndOfBibitem
\bibitem[Hunter and Weeks(2012)]{hunter12rpp}
G.~L. Hunter and E.~R. Weeks, \emph{Rep. Prog. Phys.}, 2012, \textbf{75},
  066501\relax
\mciteBstWouldAddEndPuncttrue
\mciteSetBstMidEndSepPunct{\mcitedefaultmidpunct}
{\mcitedefaultendpunct}{\mcitedefaultseppunct}\relax
\EndOfBibitem
\bibitem[Bernal(1964)]{bernal64}
J.~D. Bernal, \emph{Proc. Roy. Soc. London. Series A}, 1964, \textbf{280},
  299--322\relax
\mciteBstWouldAddEndPuncttrue
\mciteSetBstMidEndSepPunct{\mcitedefaultmidpunct}
{\mcitedefaultendpunct}{\mcitedefaultseppunct}\relax
\EndOfBibitem
\bibitem[Mason \emph{et~al.}(1995)Mason, Bibette, and Weitz]{mason95emul}
T.~G. Mason, J.~Bibette and D.~A. Weitz, \emph{Phys. Rev. Lett.}, 1995,
  \textbf{75}, 2051--2054\relax
\mciteBstWouldAddEndPuncttrue
\mciteSetBstMidEndSepPunct{\mcitedefaultmidpunct}
{\mcitedefaultendpunct}{\mcitedefaultseppunct}\relax
\EndOfBibitem
\bibitem[Ikeda \emph{et~al.}(2012)Ikeda, Berthier, and Sollich]{ikeda12}
A.~Ikeda, L.~Berthier and P.~Sollich, \emph{Phys. Rev. Lett.}, 2012,
  \textbf{109}, 018301\relax
\mciteBstWouldAddEndPuncttrue
\mciteSetBstMidEndSepPunct{\mcitedefaultmidpunct}
{\mcitedefaultendpunct}{\mcitedefaultseppunct}\relax
\EndOfBibitem
\bibitem[Scheffold \emph{et~al.}(2013)Scheffold, Cardinaux, and
  Mason]{scheffold13}
F.~Scheffold, F.~Cardinaux and T.~G. Mason, \emph{Journal of Physics: Condensed
  Matter}, 2013, \textbf{25}, 502101\relax
\mciteBstWouldAddEndPuncttrue
\mciteSetBstMidEndSepPunct{\mcitedefaultmidpunct}
{\mcitedefaultendpunct}{\mcitedefaultseppunct}\relax
\EndOfBibitem
\bibitem[Paredes \emph{et~al.}(2013)Paredes, Michels, and Bonn]{dinkgreve13}
J.~Paredes, M.~A.~J. Michels and D.~Bonn, \emph{Phys. Rev. Lett.}, 2013,
  \textbf{111}, 015701\relax
\mciteBstWouldAddEndPuncttrue
\mciteSetBstMidEndSepPunct{\mcitedefaultmidpunct}
{\mcitedefaultendpunct}{\mcitedefaultseppunct}\relax
\EndOfBibitem
\bibitem[Dinkgreve \emph{et~al.}(2015)Dinkgreve, Paredes, Michels, and
  Bonn]{dinkgreve15}
M.~Dinkgreve, J.~Paredes, M.~A.~J. Michels and D.~Bonn, \emph{Phys. Rev. E},
  2015, \textbf{92}, 012305\relax
\mciteBstWouldAddEndPuncttrue
\mciteSetBstMidEndSepPunct{\mcitedefaultmidpunct}
{\mcitedefaultendpunct}{\mcitedefaultseppunct}\relax
\EndOfBibitem
\bibitem[Dinkgreve \emph{et~al.}(2018)Dinkgreve, Michels, Mason, and
  Bonn]{dinkgreve18}
M.~Dinkgreve, M.~A.~J. Michels, T.~G. Mason and D.~Bonn, \emph{Phys. Rev.
  Lett.}, 2018, \textbf{121}, 228001\relax
\mciteBstWouldAddEndPuncttrue
\mciteSetBstMidEndSepPunct{\mcitedefaultmidpunct}
{\mcitedefaultendpunct}{\mcitedefaultseppunct}\relax
\EndOfBibitem
\bibitem[Basu \emph{et~al.}(2014)Basu, Xu, Still, Arratia, Zhang, Nordstrom,
  Rieser, Gollub, Durian, and Yodh]{basu14}
A.~Basu, Y.~Xu, T.~Still, P.~E. Arratia, Z.~Zhang, K.~N. Nordstrom, J.~M.
  Rieser, J.~P. Gollub, D.~J. Durian and A.~G. Yodh, \emph{Soft Matter}, 2014,
  \textbf{10}, 3027--3035\relax
\mciteBstWouldAddEndPuncttrue
\mciteSetBstMidEndSepPunct{\mcitedefaultmidpunct}
{\mcitedefaultendpunct}{\mcitedefaultseppunct}\relax
\EndOfBibitem
\bibitem[Wang and Brady(2015)]{mu15}
M.~Wang and J.~F. Brady, \emph{Phys. Rev. Lett.}, 2015, \textbf{115},
  158301\relax
\mciteBstWouldAddEndPuncttrue
\mciteSetBstMidEndSepPunct{\mcitedefaultmidpunct}
{\mcitedefaultendpunct}{\mcitedefaultseppunct}\relax
\EndOfBibitem
\bibitem[Nordstrom \emph{et~al.}(2010)Nordstrom, Verneuil, Arratia, Basu,
  Zhang, Yodh, Gollub, and Durian]{nordstrom10}
K.~N. Nordstrom, E.~Verneuil, P.~E. Arratia, A.~Basu, Z.~Zhang, A.~G. Yodh,
  J.~P. Gollub and D.~J. Durian, \emph{Phys. Rev. Lett.}, 2010, \textbf{105},
  175701\relax
\mciteBstWouldAddEndPuncttrue
\mciteSetBstMidEndSepPunct{\mcitedefaultmidpunct}
{\mcitedefaultendpunct}{\mcitedefaultseppunct}\relax
\EndOfBibitem
\bibitem[Petekidis \emph{et~al.}(2004)Petekidis, Vlassopoulos, and
  Pusey]{pusey04}
G.~Petekidis, D.~Vlassopoulos and P.~N. Pusey, \emph{Journal of Physics:
  Condensed Matter}, 2004, \textbf{16}, S3955--S3963\relax
\mciteBstWouldAddEndPuncttrue
\mciteSetBstMidEndSepPunct{\mcitedefaultmidpunct}
{\mcitedefaultendpunct}{\mcitedefaultseppunct}\relax
\EndOfBibitem
\bibitem[Mewis and Wagner(2012)]{mewis12}
J.~Mewis and N.~J. Wagner, \emph{Colloidal suspension rheology}, Cambridge
  University Press, 2012\relax
\mciteBstWouldAddEndPuncttrue
\mciteSetBstMidEndSepPunct{\mcitedefaultmidpunct}
{\mcitedefaultendpunct}{\mcitedefaultseppunct}\relax
\EndOfBibitem
\bibitem[Mason \emph{et~al.}(1996)Mason, Bibette, and Weitz]{mason96}
T.~G. Mason, J.~Bibette and D.~A. Weitz, \emph{J. Colloid Interface Sci.},
  1996, \textbf{179}, 439--448\relax
\mciteBstWouldAddEndPuncttrue
\mciteSetBstMidEndSepPunct{\mcitedefaultmidpunct}
{\mcitedefaultendpunct}{\mcitedefaultseppunct}\relax
\EndOfBibitem
\bibitem[van~der Wel \emph{et~al.}(2017)van~der Wel, Bhan, Verweij, Frijters,
  Gong, Hollingsworth, Sacanna, and Kraft]{casper17}
C.~van~der Wel, R.~K. Bhan, R.~W. Verweij, H.~C. Frijters, Z.~Gong, A.~D.
  Hollingsworth, S.~Sacanna and D.~J. Kraft, \emph{Langmuir}, 2017,
  \textbf{33}, 8174--8180\relax
\mciteBstWouldAddEndPuncttrue
\mciteSetBstMidEndSepPunct{\mcitedefaultmidpunct}
{\mcitedefaultendpunct}{\mcitedefaultseppunct}\relax
\EndOfBibitem
\bibitem[Bayles \emph{et~al.}(2016)Bayles, Squires, and Helgeson]{bayles16}
A.~V. Bayles, T.~M. Squires and M.~E. Helgeson, \emph{Soft Matter}, 2016,
  \textbf{12}, 2440--2452\relax
\mciteBstWouldAddEndPuncttrue
\mciteSetBstMidEndSepPunct{\mcitedefaultmidpunct}
{\mcitedefaultendpunct}{\mcitedefaultseppunct}\relax
\EndOfBibitem
\bibitem[Poon \emph{et~al.}(2012)Poon, Weeks, and Royall]{poon12}
W.~C.~K. Poon, E.~R. Weeks and C.~P. Royall, \emph{Soft Matter}, 2012,
  \textbf{8}, 21--30\relax
\mciteBstWouldAddEndPuncttrue
\mciteSetBstMidEndSepPunct{\mcitedefaultmidpunct}
{\mcitedefaultendpunct}{\mcitedefaultseppunct}\relax
\EndOfBibitem
\bibitem[Caggioni \emph{et~al.}(2020)Caggioni, Trappe, and Spicer]{marco20}
M.~Caggioni, V.~Trappe and P.~T. Spicer, \emph{Journal of Rheology}, 2020,
  \textbf{64}, 413--422\relax
\mciteBstWouldAddEndPuncttrue
\mciteSetBstMidEndSepPunct{\mcitedefaultmidpunct}
{\mcitedefaultendpunct}{\mcitedefaultseppunct}\relax
\EndOfBibitem
\bibitem[Brodu \emph{et~al.}(2015)Brodu, Dijksman, and Behringer]{brodu15}
N.~Brodu, J.~A. Dijksman and R.~P. Behringer, \emph{Nature Communications},
  2015, \textbf{6}, 6361\relax
\mciteBstWouldAddEndPuncttrue
\mciteSetBstMidEndSepPunct{\mcitedefaultmidpunct}
{\mcitedefaultendpunct}{\mcitedefaultseppunct}\relax
\EndOfBibitem
\bibitem[Bocquet \emph{et~al.}(2009)Bocquet, Colin, and Ajdari]{bocquet09}
L.~Bocquet, A.~Colin and A.~Ajdari, \emph{Phys. Rev. Lett.}, 2009,
  \textbf{103}, 036001\relax
\mciteBstWouldAddEndPuncttrue
\mciteSetBstMidEndSepPunct{\mcitedefaultmidpunct}
{\mcitedefaultendpunct}{\mcitedefaultseppunct}\relax
\EndOfBibitem
\bibitem[Mansard \emph{et~al.}(2013)Mansard, Colin, Chaudhuri, and
  Bocquet]{mansard13}
V.~Mansard, A.~Colin, P.~Chaudhuri and L.~Bocquet, \emph{Soft Matter}, 2013,
  \textbf{9}, 7489+\relax
\mciteBstWouldAddEndPuncttrue
\mciteSetBstMidEndSepPunct{\mcitedefaultmidpunct}
{\mcitedefaultendpunct}{\mcitedefaultseppunct}\relax
\EndOfBibitem
\bibitem[Denkov \emph{et~al.}(2009)Denkov, Tcholakova, Golemanov,
  Ananthpadmanabhan, and Lips]{denkov09}
N.~D. Denkov, S.~Tcholakova, K.~Golemanov, K.~P. Ananthpadmanabhan and A.~Lips,
  \emph{Soft Matter}, 2009, \textbf{5}, 3389--3408\relax
\mciteBstWouldAddEndPuncttrue
\mciteSetBstMidEndSepPunct{\mcitedefaultmidpunct}
{\mcitedefaultendpunct}{\mcitedefaultseppunct}\relax
\EndOfBibitem
\bibitem[Kobelev and Schweizer(2005)]{ken05}
V.~Kobelev and K.~S. Schweizer, \emph{Phys. Rev. E}, 2005, \textbf{71},
  021401\relax
\mciteBstWouldAddEndPuncttrue
\mciteSetBstMidEndSepPunct{\mcitedefaultmidpunct}
{\mcitedefaultendpunct}{\mcitedefaultseppunct}\relax
\EndOfBibitem
\bibitem[Kraft \emph{et~al.}(2011)Kraft, Hilhorst, Heinen, Hoogenraad, Luigjes,
  and Kegel]{kraft11}
D.~J. Kraft, J.~Hilhorst, M.~A.~P. Heinen, M.~J. Hoogenraad, B.~Luigjes and
  W.~K. Kegel, \emph{The Journal of Physical Chemistry B}, 2011, \textbf{115},
  7175--7181\relax
\mciteBstWouldAddEndPuncttrue
\mciteSetBstMidEndSepPunct{\mcitedefaultmidpunct}
{\mcitedefaultendpunct}{\mcitedefaultseppunct}\relax
\EndOfBibitem
\bibitem[Jones(2002)]{jones02}
R.~A.~L. Jones, \emph{Soft Condensed Matter (Oxford Master Series in Condensed
  Matter Physics, Vol. 6)}, Oxford University Press, 1st edn, 2002\relax
\mciteBstWouldAddEndPuncttrue
\mciteSetBstMidEndSepPunct{\mcitedefaultmidpunct}
{\mcitedefaultendpunct}{\mcitedefaultseppunct}\relax
\EndOfBibitem
\bibitem[Princen(1983)]{princen83}
H.~M. Princen, \emph{J. Colloid Interface Sci.}, 1983, \textbf{91},
  160--175\relax
\mciteBstWouldAddEndPuncttrue
\mciteSetBstMidEndSepPunct{\mcitedefaultmidpunct}
{\mcitedefaultendpunct}{\mcitedefaultseppunct}\relax
\EndOfBibitem
\bibitem[Trappe and Weitz(2000)]{trappe00}
V.~Trappe and D.~A. Weitz, \emph{Phys. Rev. Lett.}, 2000, \textbf{85},
  449--452\relax
\mciteBstWouldAddEndPuncttrue
\mciteSetBstMidEndSepPunct{\mcitedefaultmidpunct}
{\mcitedefaultendpunct}{\mcitedefaultseppunct}\relax
\EndOfBibitem
\bibitem[Olsson and Teitel(2007)]{teitel07}
P.~Olsson and S.~Teitel, \emph{Phys. Rev. Lett.}, 2007, \textbf{99},
  178001\relax
\mciteBstWouldAddEndPuncttrue
\mciteSetBstMidEndSepPunct{\mcitedefaultmidpunct}
{\mcitedefaultendpunct}{\mcitedefaultseppunct}\relax
\EndOfBibitem
\bibitem[Olsson and Teitel(2011)]{olsson11}
P.~Olsson and S.~Teitel, \emph{Phys. Rev. E}, 2011, \textbf{83}, 030302\relax
\mciteBstWouldAddEndPuncttrue
\mciteSetBstMidEndSepPunct{\mcitedefaultmidpunct}
{\mcitedefaultendpunct}{\mcitedefaultseppunct}\relax
\EndOfBibitem
\bibitem[St.~John \emph{et~al.}(2007)St.~John, Breedveld, and Lyon]{stjohn07}
A.~N. St.~John, V.~Breedveld and L.~A. Lyon, \emph{J. Phys. Chem. B}, 2007,
  \textbf{111}, 7796--7801\relax
\mciteBstWouldAddEndPuncttrue
\mciteSetBstMidEndSepPunct{\mcitedefaultmidpunct}
{\mcitedefaultendpunct}{\mcitedefaultseppunct}\relax
\EndOfBibitem
\bibitem[Lyon and Fernandez-Nieves(2012)]{lyon12}
L.~A. Lyon and A.~Fernandez-Nieves, \emph{Ann. Rev. Phys. Chem.}, 2012,
  \textbf{63}, 25--43\relax
\mciteBstWouldAddEndPuncttrue
\mciteSetBstMidEndSepPunct{\mcitedefaultmidpunct}
{\mcitedefaultendpunct}{\mcitedefaultseppunct}\relax
\EndOfBibitem
\end{mcitethebibliography}
\bibliographystyle{rsc} %the RSC's .bst file

\end{document}